# First Principles Study of Intrinsic and Extrinsic Point Defects in Monolayer $WSe_2$

Yujie Zheng[1,2] and Su Ying Quek[1,2,*]

[1]Department of Physics, National University of Singapore, 2 Science Drive 3, 117542, Singapore

[2]Centre for Advanced 2D Materials, National University of Singapore, Block S14, Level 6, 6 Science Drive 2, 117546, Singapore

* To whom correspondence should be addressed: phyqsy@nus.edu.sg

**Abstract**

We present a detailed first principles density functional theory study of intrinsic and extrinsic point defects in monolayer (ML) $WSe_2$. Among the intrinsic point defects, Se vacancies ($Se_{vac}$) have the lowest formation energy (disregarding Se adatoms that can be removed with annealing). The defects with the next smallest formation energies (at least 1 eV larger) are $Se_W$ (Se substituting W atoms in an antisite defect), $W_{vac}$ (W vacancies) and $2Se_{vac}$ (Se divacancies). All these intrinsic defects have gap states that are not spin-polarized. The presence of a graphite substrate does not change the formation energies of these defects significantly. For the extrinsic point defects, we focus on O, $O_2$, H, $H_2$ and C interacting with perfect $WSe_2$ and its intrinsic point defects. The preferred binding site in perfect $WSe_2$ is the interstitial site for atomic O, H and C. These interstitial defects have no gap states. The gap states of the intrinsic defects are modified by interaction with O, $O_2$, H, $H_2$ and C. In particular, the gap states of $Se_{vac}$ and $2Se_{vac}$ are completely removed by interaction with O and $O_2$. This is consistent with the significantly larger stability of O-related defects compared to H- and C-related defects. The preferred binding site for O is $Se_{vac}$, while that for H is $Se_W$. H bonded to $Se_W$ results in spin-polarized gap states, which may be useful in defect engineering for spintronics applications. The charge transition levels and ionization energies of these defects are also computed. H in the interstitial site is an effective donor, while all the other defects are deep donors or acceptors in isolated $WSe_2$ ML.

# I. INTRODUCTION

The field of two-dimensional (2D) materials has expanded rapidly since the first isolation of graphene [1-5]. It is now experimentally possible to synthesize monolayers of a wide range of 2D materials, using mechanical and chemical exfoliation, chemical vapor deposition (CVD) or molecular beam epitaxy techniques [2-5]. These monolayers are of interest in a wide range of applications, such as in flexible electronics, photonics, optoelectronics, catalysis and sensing [2-5]. To this end, a thorough fundamental understanding of the physical and chemical properties of these 2D materials is critical. Despite rapid experimental advancements, synthesizing defect-free 2D materials is still far from reach. Compared to bulk materials, defects in 2D materials are expected to have a more significant impact on their electronic properties, especially for semiconducting 2D materials, where defects may introduce states in the band gap [6-8] or serve as dopants [9,10]. The high surface-to-volume ratio in 2D materials also makes them more susceptible to unintentional reactions with chemicals in the environment, giving rise to extrinsic defects. For example, gases may interact with intrinsic defects in 2D materials, such as vacancy sites, changing their optical properties [11,12]. Some of the most interesting properties of 2D materials arise from the presence of defects. For example, single photon emission has recently been observed in 2D materials such as $WSe_2$ [13-17] and hexagonal boron nitride [18-20], and the single photons are believed to come from localized defect states, the nature of which were largely unknown.

Transition metal dichalcogenides (TMDs) are among the most common of the semiconducting 2D materials. $MoS_2$ is a prototypical TMD material that has been widely studied for applications in next-generation electronics [21-24]. Consistent with first principles predictions [7,8], sulfur vacancies ($S_{vac}$) are the most commonly observed defects in $MoS_2$ [25-29]. These $S_{vac}$ defects were found experimentally to play a large role in reducing the mobility of MoS$_2$ [28,30]. On the other hand, the threshold voltage of MoS$_2$ field-effect transistors can be tuned by controlling the density of S$_{vac}$ through surface treatments [31]. S$_{vac}$ are predicted to be deep acceptors in isolated MoS$_2$ monolayers [7]. Recent ionization dynamics experiments have found that S$_{vac}$ in MoS$_2$ monolayers on Au substrates are neutral or negatively charged [32]. The negative charge in these S$_{vac}$ comes not from MoS$_2$, but from the Au substrates, and results in the pinning of the Fermi level at the S$_{vac}$ gap states near the conduction band minimum (CBM), thus explaining the typical n-type characteristics of supported MoS$_2$ monolayers [32]. In contrast to MoS$_2$, WSe$_2$ was the first monolayer TMD discovered to have ambipolar and p-type characteristics [33-36]. WSe$_2$ has also attracted much recent attention because of the discovery of single photon emission from WSe$_2$ monolayers, [13-17] which arise from localized electronic states [37,38]. Thus far, single photon emission has not been reported in MoS$_2$. Taken together, these distinct properties of WSe$_2$ found experimentally underscore the importance of a systematic study of defects in this material.

Point defects are the most common and simple defect structures in TMDs. First principles density functional theory (DFT) studies on intrinsic point defects in TMDs have provided much useful information on the formation energies and possible charge states of these defects [6-8]. The prediction that S$_{vac}$ is the most likely point defect in MoS$_2$ has been verified experimentally, by a combination of scanning transmission electron microscopy (STEM) [25,27] and scanning tunneling microscopy (STM) [26,29]. Researchers have also enhanced the mobility of MoS$_2$ by using thiol chemistry to heal the S$_{vac}$ [28], thus indirectly showing that S$_{vac}$ are prevalent in MoS$_2$ samples. However, even though Se vacancies (Se$_{vac}$) were predicted to be the most likely point defect in WSe$_2$ [8], it was recently suggested based on scanning tunneling microscopy (STM) images, that W vacancies (W$_{vac}$) are instead the most abundant point defect in WSe$_2$ monolayers on graphite [39]. Using a combination of DFT calculations as well as STM and scanning transmission electron microscopy (STEM) experiments, we have shown that $W_{vac}$ are not present in $WSe_2$. The most abundant point defect in $WSe_2$ monolayers is found to be an extrinsic defect, with an O atom substituting a Se atom ($O_{Se}$).[40] These $O_{Se}$ defects are very stable and arise from the dissociation of $O_2$ at $Se_{vac}$ sites, which can take place easily at room temperature [40]. As discussed above, extrinsic defects can play a significant role in 2D materials because of the high surface-to-volume ratio, increasing the possibility of intrinsic defects reacting with gases and chemicals in the environment. Several extrinsic defects in 2D TMDs, such as naturally found Re impurities in $MoS_2$ [41] and Cr impurities in $WS_2$ [42], have also been discussed in the literature. Besides metal impurities and O-related defects, it is also important to consider H- and C-related extrinsic defects, because of the presence of hydrocarbons and reducing agents such as $H_2$ gas during the CVD growth process [34,43-46].

In this work, we explore, using first principles DFT calculations, the atomic and electronic structure, and the formation energies of intrinsic point defects found in monolayer $WSe_2$ supported on graphite substrates, as well as O, C and H-related extrinsic point defects in monolayer $WSe_2$. The ionization energies (charge states) of the most stable point defects in each category are also studied. $Se_{vac}$ has the lowest formation energy among all the intrinsic point defects considered. The $Se_{vac}$ gap states are passivated by O and $O_2$. The interstitial (hollow) site is the favored binding site for atomic O, H and C in perfect $WSe_2$ ML, and these defects have no gap states. $H_{ins}$ (H at the interstitial site) acts as a donor. H adsorbs stably at $W_{vac}$ and $Se_W$ (antisite defect with Se substituting a W site), giving rise to defects with spin-polarized gap states. Thus, interesting magnetic properties may be induced if these H-related defects can be engineered, e.g., by exposing $WSe_2$ ML to atomic H [47]. Besides $H_{ins}$, all the other defects considered here have ionization energies that are larger than 0.6 eV, indicating that the defects are not effective dopants in isolated $WSe_2$ monolayers. However, $W_{vac}$ and $Se_W$ may become acceptors if the Fermi level ($E_F$) is positioned close to mid-gap by a substrate.

## II. METHODS

The DFT calculations were performed using projector augmented wave (PAW) [48] potentials with the generalized gradient approximation GGA-PBE [49] for the exchange-correlation functional, as implemented in the VASP code [50,51]. Van der Waals interactions were included by adding Grimme's D2 corrections [52]. Spin-polarization was included for all neutral defect structures and for $O_2$. The atoms are all relaxed with a force convergence criteria of 0.01 eV/Å for $WSe_2$ monolayers (ML), and 0.05 eV/Å for the $WSe_2$ ML supported on graphite. A kinetic energy cutoff of 400 eV is used for the plane wave basis set. We have tested that the total energy of bulk $WSe_2$ is converged for this value of the energy cutoff, as well as with a Monkhorst Pack k-grid sampling of 10 x 10 x 4 in the bulk $WSe_2$ unit cell. The relaxed lattice parameters for bulk $WSe_2$ compare well to the experimental values (theory: a = 3.327 Å, c = 12.788 Å; experiment [53]: a = 3.282 Å, c = 12.96 Å). The formation energies and densities of states (DOS) of defects in isolated $WSe_2$ ML were obtained with a 5 x 5 supercell with a vacuum height of larger than 20 Å to separate periodic slabs. A 2 x 2 k-mesh was used for geometry optimization in this 5 x 5 supercell, while a 6 x 6 k-mesh was used for the DOS calculations. The graphite lattice constant obtained theoretically is also close to experiment (theory: a = 2.464 Å, c = 6.432 Å; experiment [54]: a = 2.461 Å, c = 6.709 Å), with a converged k-mesh of 16 x 16 x 6 in the graphite unit cell. $WSe_2$ ML on graphite was modeled using 3 layers of graphite, with a 3 x 3 $WSe_2$ supercell on top of a 4 x 4 supercell of graphite and 13 Å of vacuum. In this supercell, the strain on $WSe_2$ was 0.61 % and the strain on graphite was -0.65 %. For this supercell, we used a k-mesh of 6 x 6 x 1 for geometry optimization and 10 x 10 x 1 for DOS calculations. Defects in $WSe_2$ ML on graphite were studied with an even larger cell, consisting of a 2×√3R30° supercell of the WSe$_2$/graphite supercell described above. Here, we used a k-mesh of 2 x 2 x 1 for geometry optimization and 4 x 4 x 1 for DOS calculations.

The formation energies $E_f(x,q)$ for the defects ($x$) in charge state $q$ are defined as [55]

$$E_f(x,q) = E(x,q) - E_{pristine} - \sum_i n_i \mu_i + q(E_v + E_F) \tag{1}$$

where $E(x,q)$ and $E_{pristine}$ are the total energies of the WSe$_2$/graphite supercell with and without the defect, respectively. $n_i$ denotes the number of atoms of element $i$ that have been added ($n_i > 0$) or removed ($n_i < 0$), and $\mu_i$ is the chemical potential of element $i$. $E_F$ is the Fermi energy relative to valance band maximum $E_v$ of WSe$_2$ ($E_F$ values range from 0 to the size of the band gap). We elaborate on the determination of $E_v$ below.

The chemical potentials of W and Se are linked by the stability of WSe$_2$, *i.e.*

$$\mu_{WSe_2} = \mu_W + 2\mu_{Se} \tag{2}$$

$\mu_{WSe_2}$ is the total energy per formula unit of bulk WSe$_2$ (the conclusions are unchanged if we use instead the total energy per formula unit of monolayer WSe$_2$). $\mu_W^{max}$ and $\mu_{Se}^{max}$ are total energies per atom of bcc W metal and the molecular crystal ($R\bar{3}$ phase) of Se$_6$ molecules, respectively. The minimum of the Se chemical potentials can be derived from formula (2), *i.e.*, $\mu_{Se}^{min} = (\mu_{WSe_2} - \mu_W^{max})/2$. The chemical potentials of O and H are half of the total energies of O$_2$ and H$_2$ gas, respectively. The chemical potential of C is the total energy per atom of graphite. For charge neutral intrinsic point defects, equation (1) reduces to

$$E_f(x) = E(x) - E_{pristine} - \sum_i n_i \mu_i \tag{3}$$

For charged defects, the transition energy is defined as [55]

$$\varepsilon(q/q') + E_v = [E(x,q) - E(x,q')]/(q'-q) \tag{4}$$

where $q$ and $q'$ are two different charge states of the same defect ($x$). However, for charged defects in a 2D system, the Coulomb interaction energy between a charged defect and its periodic images diverges as $L_z \to \infty$ [55]. The ionization energies (IE) of a defect are special cases of the defect transition energies, *i.e* $\varepsilon(q,q')$ relative to the band edges [55]. $\varepsilon(+/0)$ relative to conduction band minimum (CBM) defines a donor IE, and $\varepsilon(0/-)$ relative to valance band maximum (VBM), defines an acceptor IE [55]. The IE for a defect in a 2D material with a supercell of x-y area $S$ and height $L_z$ takes the asymptotic form [55]

$$IE(S, L_z) = IE_0 + \frac{\alpha}{\sqrt{S}} + \frac{\beta}{S} L_z \tag{5}$$

where $IE(S, L_z)$ is a supercell-size-dependent quantity and $IE_0$ (converged) is the true and size-independent *IE*. $\alpha$ and $\beta$ are constants for each defect. According to equation (5), finding $IE(S, L_z)$ at different $S$ and $L_z$ will allow us to find $IE_0$ in 2D systems. If the supercell size is large enough, $IE(S, L_z)$ at a given fixed $S$ should be linear with respect to $L_z$. The intercept at $L_z = 0$, $IE_0 + \frac{\alpha}{\sqrt{S}}$, is found for each $S$. This intercept should be linear with respect to $\frac{1}{\sqrt{S}}$, and the intercept in this plot will be $IE_0$. We use this method (Method 1) with the dimensions of the supercell varying from 20 to 40 Å for $L_z$ and 7 × 7 to 9 × 9 supercells for $S$. A 3 × 3 × 1 k-grid is used for these calculations.

The defect itself affects the band structure including the value $E_v$ in the calculation [56]. In order to obtain the value that $E_v$ would be for a larger supercell, a two-step procedure was used. First, $E_v$ is calculated for the perfect isolated WSe$_2$ ML by performing a band-structure calculation at the K point for each $L_z$. Next, the 1s core level of the perfect isolated WSe$_2$ ML is aligned with the 1s core level of W in the supercell, furthest from the defect site; this alignment procedure is used to give the value of $E_v$ in the supercell.

Although the above method can calculate the IEs correctly, it requires many calculations. Komsa *et al.* [57] reported that different types of defects in a given 2D material have similar errors in the IE for fixed cell size and shape. Thus, using Method 1 as a benchmark, these errors can be estimated. We have found that a "special" cell ($9 \times 9$ supercell with a vacuum height of 30 Å, corresponding to $L_z = 33.4$ Å) gives IEs that are within 0.2 eV of the converged $IE_0$ obtained using Method 1. Using this "special" cell, we thus estimate the IE of a larger set of defects. We call this Method 2.

### III. RESULTS AND DISCUSSION

#### A. Intrinsic defects

##### 1. Atomic structure and Formation Energies

Graphite is a flat conducting substrate that is used for direct CVD growth of $WSe_2$ ML and for characterization of $WSe_2$ ML in STM [40,58]. Haldar et al. has studied in detail several intrinsic defects in isolated $WSe_2$ ML, in particular Se and W single and double vacancies, as well as interstitial defects [8]. Here, we study these and other intrinsic defects in $WSe_2$ ML on graphite. Fig. 1 shows the side view of the 13 different intrinsic point defects considered for $WSe_2$ ML on graphite. Se atoms in the upper and lower planes are labeled Se1 and Se2, respectively. $Se1_{vac}$ (Fig. 1a) refers to a defect where Se1 is missing, and $Se2_{vac}$ (Fig. 1b) refers to that where Se2 is missing. $W_{vac}$ (Fig. 1c) refers to a W vacancy, and $2Se_{vac}$ (Fig. 1d) refers to a Se divacancy, where both Se1 and Se2 are missing (one on top of the other). In addition to vacancy sites, we have also considered antisite defects, adatoms, and interstitial defects where the Se and W atoms are in between the $WSe_2$ ML and the graphite substrate (Fig. 1l: $Se_{int}$; Fig. 1m: $W_{int}$). Se adatom and W adatom defects are shown in Fig. 1j ($Se_{ad}$) and 1k ($W_{ad}$), respectively. For the antisite defects, the nomenclature is as follows. $X_Y$ implies that element X has substituted element Y. $2Se_W$ refers to two Se atoms substituting a W atom, while $W_{2Se}$ refers to one W atom substituting two Se atoms. $Se_W$ (Fig. 1e), $W_{Se1}$ (Fig. 1f), $W_{Se2}$ (Fig. 1g), $2Se_W$ (Fig. 1h) and $W_{2Se}$ (Fig. 1i) are considered here. Of these 13 defects, only $2Se_W$ (Fig. 1h), $W_{2Se}$ (Fig. 1i), $W_{ad}$ (Fig. 1k), $Se_{int}$ (Fig. 1l) and $W_{int}$ (Fig. 1m) lead to significant changes in the atomic positions of WSe$_2$. In the $W_{vac}$ structure, the Se-W bonds next to the defect site shorten by 1.6%, while in the $Se_{vac}$ and $2Se_{vac}$ structures, the adjacent Se-W bnods lengthen by 0.4%. In $Se_W$, the Se atom sits in the three-fold site occupied by W, forming three Se-Se bonds of equal bond length, 3.2% larger than a Se-W bond in perfect WSe$_2$ ML. Interestingly, the W adatom drops into the WSe$_2$ layer and bonds with a W atom and three neighboring Se atoms (Fig. 1k).

The formation energies of these intrinsic defects are shown in Fig. 2a for a range of chemical potentials ranging from the W-rich to the Se-rich limits. $Se1_{vac}$ and $Se2_{vac}$ have very similar formation energies and DOS. Thus, henceforth, we shall refer to them both as $Se_{vac}$. $Se_{ad}$ has the lowest formation energy regardless of the chemical potential. However, the formation energy is still positive, which indicates that the Se adatom can desorb or diffuse away during the high temperatures of CVD growth or subsequent annealing processes. Similarly, the interstitial Se atom between WSe$_2$ and graphite ($Se_{int}$) may also diffuse away. However, the vacancies and antisite defects, once formed, are less likely to be removed. Of all the vacancies and antisite defects, $Se_{vac}$ has the lowest formation energy even under Se-rich conditions. The formation energy of $Se_{vac}$ ranges from 2.2 eV (W-rich) to 2.8 eV (Se-rich). The defects with the next smallest formation energies are Se divacancy $2Se_{vac}$, $Se_W$ antisite defect, and $W_{vac}$, which all have very similar ranges of formation energies, between ~3.7-3.9 eV and ~5.1-5.5 eV (Fig. 2a; Table I). These formation energies are significantly larger than that of $Se_{vac}$, indicating that $Se_{vac}$ is by far the most abundant intrinsic point defect in WSe$_2$ monolayers on graphite. Of the three defects, 2Se$_{vac}$, Se$_W$ and W$_{vac}$, Se$_W$ has the lowest formation energy under Se-rich conditions, while 2Se$_{vac}$ has the lowest formation energy under W-rich conditions. Most CVD growth conditions take place in the Se-rich limit, due to the low sublimation temperature of Se, and the abundance of Se vapor.[34,36,43,59] In this limit, our calculations predict that the most likely intrinsic defects for WSe$_2$ ML on graphite are Se$_{vac}$, followed by Se$_W$, W$_{vac}$ and 2Se$_{vac}$ (disregarding adatoms and interstitials), in that order. To investigate the effect of the graphite substrate on the formation energies, we also compute the formation energies of selected intrinsic defects in isolated WSe$_2$ ML (Fig. 2b; Table I). In general, the energy differences caused by the graphite substrate are small, compared to the actual formation energies of the defects. The relative formation energies are also unchanged by the graphite substrate. STEM experiments [40] have identified Se$_{vac}$ to be the most abundant defect in WSe$_2$ ML. Se$_W$ and 2Se$_{vac}$ were also detected, but interestingly, no W$_{vac}$ were found. It is possible that the 2Se$_{vac}$ defects were created due to electron bombardment during the STEM imaging process. We note that the electron beam could also knock away O atoms that we predict to be substituted in the Se$_{vac}$ sites [40].

2. Electronic Structure

The DOS projected onto the defect sites are plotted in Fig. 3 for the intrinsic defects in WSe$_2$ ML on graphite. Except for Se$_{ad}$, all the other intrinsic defects here have defect states in the band gap of WSe$_2$. The absence of gap states in Se$_{ad}$ is consistent with the fact that it is the most energetically favorable defect. Spin polarized calculations were performed for these intrinsic defects in isolated WSe$_2$. None of these defects is spin polarized, except for W$_{Se}$. Since W$_{Se}$ has a very large formation energy, the details are omitted here.

Two commonly used experimental methods to identify defects in 2D materials are STM and STEM studies. STEM is sensitive to relative atomic numbers in the sample, the intensity of the image being approximately proportional to the square of the atomic number. However, it will be difficult to detect atoms with much smaller atomic numbers relative to the other atoms in the sample, and damage to the sample can occur from the transmitting electron beam. STM images result from the tunneling of electrons between an STM metallic tip and the sample; these electrons are not energetic enough to create defects in the sample. Within the Tersoff-Hamann approximation [60], the STM image intensity is proportional to the local density of states that are available to contribute to the tunneling current, evaluated at the tip position. The relative voltage difference between the tip and the sample in turn determine the energy range of states available to contribute to the current. Thus, STM does not image the atoms directly, but rather the electronic states in an energy range that depends on the applied bias and the Fermi level. DFT calculations of electronic states are therefore very valuable to facilitate the identification of point defects observed in STM images.

Fig. 4 shows the STM images for a perfect $WSe_2$ ML on graphite (a commonly used substrate for STM studies of 2D materials). It is interesting to note that the bright spots in these images do not necessarily correspond to the positions of the Se atoms (which are located on the top surface). Bright spots are located at the Se atoms for images at -0.8 V and 1.2 V (sample bias), which respectively probe states 0.4 eV into the valence band, and 0.1 eV into the conduction band. However, for the image at -0.6 V sample bias (states 0.2 eV into the valence band), the bright spots are at the hollow sites. For the image at 1.3 V sample bias (states 0.2 eV into the conduction band), the bright spots are located at both the hollow sites and the Se sites. Generally, STM images can also be affected by other factors such as the tip-sample distance [61,62] and tip condition.

We next simulate the STM images of all the thirteen intrinsic defects considered in this work, as shown in Fig. 5. The bias-dependent STM images for these defects are all distinct, and reflect the local density of states of the defects near the band edges. The STM image for Se$_{ad}$ has a much brighter spot at the Se adatom for both positive and negative bias voltages; this is expected because the Se adatom is much closer to the STM tip than the WSe$_2$ ML. Other atomic adsorbates would also result in very similar images. For the other point defects, it is possible that STM images can act as a "fingerprint" to help identify the point defect, but it is clear that multiple images acquired at different bias voltages for the same defect will greatly help to reduce the uncertainty in assigning the defect type. Since Se$_{vac}$ has the lowest formation energy among these defects, we discuss in particular the STM images of Se$_{vac}$. These images are similar to those predicted for S$_{vac}$ in MoS$_2$ [26], including simulations that include the STM tip explicitly [61]. The image for occupied states has a dark triangular depression that extends over several atomic positions, while that for unoccupied states has a dark circle surrounded by a bright circular shape. Experimentally, STM images for MoS$_2$ ML on graphite and on Au have uncovered point defects that resemble these images [26,29]. However, the STM images for WSe$_2$ ML on graphite contain other point-like defects that do not resemble the simulated STM images for Se$_{vac}$ [39,40]. Bias-dependent STM images for the most commonly observed point defect in the WSe$_2$ ML shows that the occupied states have a gray triangular area or a dark three-point star shape surrounded by a bright triangle [39,40]. On the other hand, the image corresponding to the unoccupied states has dark depressions but does not have the bright circle that is seen in the Se$_{vac}$ image [39,40]. We show in Reference [40] that the most commonly observed defect observed in these STM images corresponds to Se$_{vac}$ passivated by a strongly bound O atom (O$_{Se}$).

### B. Extrinsic defects

#### 1. Atomic structure and Formation Energies

Extrinsic defects arising from contaminant transition metal atoms have been studied in the past for TMD ML [63,64]. Here, we focus on extrinsic point defects involving elements that are present in a typical CVD growth process, i.e. O, H and C. O-related defects can also be formed due to exposure to $O_2$ in ambient conditions. We consider O, $O_2$, H, $H_2$ and C atoms interacting with perfect $WSe_2$ ML as well as with the more stable intrinsic point defects ($Se_{vac}$, $Se_W$, $W_{vac}$ and $2Se_{vac}$). The relaxed structures of the O-, H-, and C-related point defects are shown in Figs. 6, 7 and 8, respectively, and the formation energies are given in Table II, together with the bond lengths of $O_2$ and $H_2$ where applicable. The nomenclature of defects follows the following conventions: (a) ad: adatom; (b) ins: insertion; (c) $X_{Se/W}$: X on $Se/W_{vac}$; (d) X-Y: X adsorbed on intrinsic defect Y. The formation energy is defined in two ways:

$$E_{f1} = E_{defect} - E_{pristine} - \sum_i n_i \mu_i \qquad (6)$$

where $E_{defect}$ and $E_{pristine}$ are the total energy with and without defects, respectively.

$$E_{f2} = E_{defect} - E_{intrinsic-defect} - \sum_i n_i \mu_i \qquad (7)$$

where $E_{defect}$ and $E_{intrinsic-defect}$ are the total energy of the supercell with the intrinsic defect, with and without impurity atom, respectively. In both formula, $n_i$ indicates the number of atoms of element $i$ and $\mu_i$ is the corresponding chemical potential. The chemical potentials of W and Se are those corresponding to Se-rich conditions. The chemical potentials of O, H and C are the total energy per atom of $O_2$, $H_2$ and graphite, respectively. The formation energies relative to atomic O, H and C are also indicated in brackets for selected defects, computed using the experimental gas phase bond dissociation energies of $O_2$ and $H_2$ [65] and the computed binding energy per C atom in graphite. The computed bond dissociation energy for $H_2$ is 4.5 eV (same as the experimental value), but that for $O_2$ is different (see Table III below). $E_{f1}$ depends on the chemical potential of Se and W if an intrinsic defect is involved.

$E_{f1}$ can be regarded as the formation energy of extrinsic point defects during the CVD growth, whereas $E_{f2}$ can be regarded as the formation energy of the extrinsic point defect after CVD growth, and is generally much smaller (more negative) than the corresponding value of $E_{f1}$. On the other hand, the formation energies given relative to atomic O, H or C are a better estimate for the stability of the isolated O, H or C atom related extrinsic defects after formation. This is because there is usually no atomic O, H or C nearby to recombine with the O, H or C that is released from the defect site, which would then be released as atoms initially. Zero point energies are not included in the formation energies in Table II. We computed zero point energy corrections to the formation energies for $O_2$-$Se_{vac}$ and $H_2$-$Se_{vac}$, and found that the formation energies changed from -2.80 eV to -2.72 eV for $O_2$-$Se_{vac}$ and from -0.77 eV to -0.64 eV for $H_2$-$Se_{vac}$. These corrections are quite small and will not change the conclusions of the manuscript. Formation energies as defined above will take on a more negative value for more stable defects, and a positive value for unstable defects. For convenience, we also use the term binding energy below, for the binding of impurity atoms to the $WSe_2$ lattice or intrinsic defect sites. The binding energy has the opposite sign as the formation energy (the more positive the binding energy, the stronger the impurity binds).

*(a) O-related defects.* The interaction of oxygen with 2D materials is an important problem because oxidation is often the cause of material degradation in these systems. For example, the dissociative chemisorption of $O_2$ on phosphorene leads to its decomposition [66]. On the other hand, the reversible passivation of sulphur vacancies in $MoS_2$ with $O_2$ can tune its optical properties [12]. Fig. 6 shows the relaxed atomic structures of the O-related point defects in $WSe_2$ ML. Except for O-Sew (O adsorbed at the Sew defect; Fig. 6h) where the Se atom at the antisite is pushed away from O, O and $O_2$ do not induce significant distortion of the lattice (Fig. 6). The O atom can form three bonds with W atoms at $Se_{vac}$ (Fig.

6a), while it forms two bonds to Se in O-Se$_W$ and O$_W$ defects. The oxygen adatom forms one bond to the Se atom. At O$_{ins}$, O sits in a hollow site. O$_2$ dissociates into two O atoms at the 2Se$_{vac}$ site (Fig. 6g), with an O-O distance of about twice the bond length in O$_2$ (Table II). The O$_2$ bond length is only slightly lengthened when adsorbed at the Se$_W$ and W$_{vac}$ sites, but is lengthened by about 20% at the Se$_{vac}$ site (Table II). It is not energetically favorable for O$_2$ to bind to W$_{vac}$ and Se$_W$. However, O$_2$ binds strongly to Se$_{vac}$ and 2Se$_{vac}$. The strength of interaction between O$_2$ and the defect sites can be directly correlated with the amount of electron transfer from the defect site to O$_2$ (Table V). There is significant electron transfer from Se$_{vac}$ and 2Se$_{vac}$ to O$_2$. The electron charge enters the antibonding orbital of O$_2$ and weakens the O=O bond, thus facilitating the dissociation of O$_2$ [40]. Similar observations have been made for O$_2$ on Au catalysts [67]. The Se$_{vac}$ is the most stable binding site for atomic O. O binds at Se$_{vac}$ to form O$_{Se}$, with a binding energy of 7.05 eV relative to atomic O. Thus, O$_{Se}$ is a very stable defect, and once formed, O is unlikely to be removed from the Se$_{vac}$ site. O can also bind stably to a W$_{vac}$ defect, forming O$_W$. However, since O$_2$ cannot bind to W$_{vac}$, O$_W$ can only form if atomic O is available from some other source. The same applies to O-Se$_W$. In perfect WSe$_2$, the most stable binding site for O corresponds to O$_{ins}$, followed by O$_{ad}$.

It is noted that DFT with standard exchange-correlation functionals, such as PBE-D2 used here, may not be able to predict quantitatively the absolute adsorption energies of O and O$_2$, because the predicted value of the bond dissociation energy of gas phase O$_2$ deviates significantly from the experimental value (Table III). However, we are more interested in qualitative results such as the relative adsorption energies, and it has been shown that both PBE and RPBE [68] functionals give qualitatively similar results for the sticking curves of O$_2$ on Al(111), even though the O$_2$ bond dissociation energy computed with RPBE is one of the closest to experiment among the gradient corrected functionals [69]. To obtain an idea of the magnitude of the error in absolute adsorption energies, we show in Table III the binding energies of $O_2$ to $Se_{vac}$ using different exchange-correlation functionals (PBE-D2, RPBE, RPBE with Grimme's D3 correction [70], and BEEF-vdW [71]). RPBE has been used successfully to describe the activation of $O_2$ on metal surfaces [69,72], RPBE-D3 reproduces the properties of liquid water [73], while BEEF-vdW was developed for surface science and catalysis studies. BE1, the binding energy relative to the energy of $O_2$, is 2.80 eV using PBE-D2, and ~1.9-2.2 eV using the other functionals. Thus, in comparison, PBE-D2 overbinds the $O_2$ molecule at the $Se_{vac}$ defect site by about 0.6-0.9 eV. If we assume conservatively that $O_2$ overbinds by 0.9 eV and O overbinds by 0.45 eV relative to the energy per atom in $O_2$, we see that all the formation energies for O-related defects in Table II are still very negative (stable), except for O-$Se_W$ which was marginally stable (with a formation energy of -0.01 eV). It is also instructive to obtain the binding energy relative to atomic O, as this is important for the O atom related defects. We also compute the binding energy, BE2, of $O_2$ on $Se_{vac}$, relative to the energy of two O atoms. However, as noted above, the bond dissociation energy of $O_2$ is overestimated with the functionals, when compared to experiment. If instead the $O_2$ bond dissociation energy from experiment is used to compute the binding energy of $O_2$ to $Se_{vac}$, relative to the energies of two O atoms (BE3), we find that BE3 corresponding to PBE-D2 is larger than BE3 from other functionals, but is close to the values of BE2 obtained by RPBE and RPBE-D3 (Table III). In Table II, we have used the $O_2$ bond dissociation energy from experiment [65] to compute the formation energies relative to atomic O.

*(b) H-related defects.* Hydrogen is a commonly found impurity in many materials, with hydrogen embrittlement [74] being a serious problem in metals, such as aluminium and titanium. Hydrogen has been used to create n-doped contacts in TMDs ($MoS_2$), by using hydrogen silsesquioxane (HSQ) as a photoresist coupled with low electron beam dosages to destroy the Si-H bonds in HSQ and release H atoms into the TMD [47,75]. $H_2$ gas is also used as a reducing agent to obtain better quality TMD films during CVD growth [34]. Table II shows that generally, the binding energies of H and $H_2$ in $WSe_2$ ML and its intrinsic defect sites are much smaller than those of O and $O_2$. Using the atomic energy in $H_2$ as a chemical potential, the formation energies $E_{f1}$ of all H-related defects considered here are positive, indicating that H-related defects are unlikely to form in $WSe_2$ during the natural CVD growth process involving $H_2$. If H atoms are available for reaction with perfect $WSe_2$ (e.g. from HSQ), the most stable defect structure is $H_{ins}$, where H is inserted into the hollow site (Table II; Fig. 7b). Similar to O-related defects, the hollow site is preferred to the Se atop site (in $O_{ad}$ and $H_{ad}$). However, the binding energy for H at $H_{ins}$ is only 0.88 eV (relative to atomic H). Thus, these H atoms can be removed by annealing. $H_{ins}$ is an effective donor in $WSe_2$ ML (Table IV below), consistent with the effectiveness of H for creating n-doped contacts in TMDs. The small binding energy for H implies that the TMD should be covered with a contact or protective layer to prevent the H dopants from being removed with annealing.

H and $H_2$ bind favorably to several of the intrinsic defect sites. H atoms bind to $Se_{vac}$, $W_{vac}$ and $Se_W$. The formation energy for H-$Se_W$ is the smallest (-0.43 eV relative to atomic energy in $H_2$), followed by $H_W$ and $H_{Se}$ (-0.27 eV relative to atomic energy in $H_2$). Interestingly, there is more lattice distortion in $H_W$ and H-$Se_W$ than in $H_{Se}$ (Fig. 7i, h, a). Yet, the formation energy for $H_{Se}$ is the largest (least stable). On the other hand, $H_2$ binds to $Se_{vac}$ with a formation energy of -0.77 eV. This implies that for the defect concentrations considered here, it is thermodynamically more favorable (by 0.23 eV) for $H_2$ to bind to a single $Se_{vac}$ site, leaving another $Se_{vac}$ empty, than it is for one H atom to bind to each $Se_{vac}$ site, forming two copies of $H_{Se}$. Examining the structure for $H_2$-$Se_{vac}$ (Fig. 7d), it can be seen that $H_2$ dissociates without any energy barrier at $Se_{vac}$ to form two H atoms which both bind at the $Se_{vac}$ site. One of the H atoms forms a single bond with a neighboring W atom, while the other forms two bonds with the other two nearest neighbor W atoms. $H_2$ has also dissociated in $H_2$-$2Se_{vac}$ (Fig. 7g), but the binding energy there is much smaller (0.39 eV). This is unlike $O_2$, which binds much more strongly to $2Se_{vac}$ than to $Se_{vac}$. The local bonding configuration of each O (H) atom in $O(H)_2$-$2Se_{vac}$ is similar to that of $O(H)_{Se}$. Thus, the difference in relative binding energies between $O_2$ and $H_2$ at $2Se_{vac}$ is consistent with the strong binding of O in $O_{Se}$ and weak binding of H in $H_{Se}$. The weak binding of H in $H_{Se}$ may in turn be related to the long H-W bond lengths in $H_{Se}$, compared to those in $H_2$-$Se_{vac}$ (H, with a very small atomic radius, favors short bonds). Similar to $O_2$, $H_2$ does not bind to $W_{vac}$. In contrast to $O_2$, it is energetically favorable for $H_2$ to bind at $Se_W$, with a formation energy of -0.67 eV. The adsorption of $H_2$ at $Se_W$ is accompanied by a significant distortion in the lattice, with one of the Se-Se bonds being broken (Fig. 7e). Likewise, a Se-Se bond is broken when H binds to $Se_W$ (its most stable binding site), where H is inserted into a Se-Se bond (Fig. 7h).

*(c) C-related defects.* None of the C-related defects studied here is thermodynamically stable when the chemical potential for C is the energy per atom in graphite. Thus, it may be less likely to form these C-related defects. However, once the defects are created, they are stable in the absence of reactants with C (Table II, formation energies using the chemical potential of atomic C). The most favorable binding site for C atoms is $Se_{vac}$ (Table II; Fig. 8a). In perfect $WSe_2$ ML, the C atom prefers the interstitial site to the Se adatom site (just like H and O). The introduction of C leads to significant lattice distortions in $C_W$ (C at a $W_{vac}$ site) and $C$-$Se_W$ (Fig. 8e-f). Otherwise, no significant lattice distortion is seen in $C_{Se}$, $C_{ins}$, $C_{ad}$ and $2C$-$2Se_{vac}$ (Fig. 8a-d).

2. Electronic structure

The spin-polarized densities of states (DOS) for O, H and C-related defects in $WSe_2$ ML are shown in Figs. 9, 10 and 11, respectively. Of all the different species interacting with intrinsic defects in WSe2 ML, we find that only O and $O_2$ interacting with $Se_{vac}$ and $2Se_{vac}$ fully passivate the gap states of the intrinsic defects ($Se_{vac}$ and $2Se_{vac}$ in this case) (Fig. 9 a, d, g). In the literature [12], it has been shown that $O_2$ passivates the gap states of $S_{vac}$ in $MoS_2$ ML, so that the photoluminescence of $MoS_2$ ML can be tuned by exposure to $O_2$ and subsequent annealing, which removes the $O_2$ from $S_{vac}$. We show in reference [40] that $O_2$ can dissociate at room temperature at $Se_{vac}$ sites in WSe2 ML, giving $O_{Se}$. This is not possible for $S_{vac}$ sites in $MoS_2$ ML [40]. The binding energy of atomic O to $Se_{vac}$ (7.05 eV) is much larger than that of $O_2$ to $S_{vac}$ (computed to be 2.02 eV with PBE-D2, with a possible overestimate of ~0.6-0.9 eV according to the analysis in Table III). Thus, O at $Se_{vac}$ is very stable and will not be removed by annealing, in contrast to $O_2$ at $S_{vac}$ sites [12].

For H, $H_2$ and C interacting with $Se_{vac}$ and $2Se_{vac}$, the gap states are modified rather than removed (Fig. 10a, f; Fig. 11a). This is consistent with the much stronger interaction between O and $O_2$ with $Se_{vac}$ and $2Se_{vac}$, compared to H, $H_2$ or C. On the other hand, despite the reasonably strong interaction between O and $W_{vac}$, $O_W$ still has gap states (Fig. 9i). The gap states in $W_{vac}$ and $Se_W$ are modified by interaction with O, $O_2$, H, $H_2$ and C. The difficulty in removing gap states for $W_{vac}$ and $Se_W$ may be related to the fact that there are multiple discrete levels in the band gap for these intrinsic defects (Fig. 3e, f). In perfect WSe2 ML, O, H and C all prefer to bind in the interstitial site. The resulting defect geometries ($O_{ins}$, $H_{ins}$ and $C_{ins}$) have no gap states (Fig. 9b, 10b, 11b). O in the adatom site ($O_{ad}$) also has no gap states, but $H_{ad}$ and $C_{ad}$ both have spin polarized gap states.

Figures 9 to 11 show that the following extrinsic point defects are spin-polarized: $O_2$-$Se_W$ (Fig. 9e), $O_2$-$W_{vac}$ (Fig. 9f), $H_{Se}$ (Fig. 10a), $H_{ad}$ (Fig. 10c), H-$Se_W$ (Fig. 10d), $H_W$ (Fig. 10e), and $C_{ad}$ (Fig. 11c). $O_2$-$Se_W$ and $O_2$-$W_{vac}$ are not stable, while $H_{ad}$ has a very small binding energy of 0.05 eV (Table II). Thus, we focus on $H_{Se}$, $H_W$ and H-$Se_W$. These three defects all have multiple spin-polarized defect states in the band gap, and may be interesting from the point of view of defect engineering for spin-related applications. Fig. 12 shows the isosurface contour plots of the difference between the spin up and spin down charge densities for these defects. All these defects have regions of net spin up charge densities that extends at least one lattice spacing around the defect site, while $H_W$ and $H_{Se}$ also have net spin down charge densities localized near the H atoms. Earlier, we had shown that it is more energetically favorable for two H atoms to bind to a single $Se_{vac}$ site, leaving one $Se_{vac}$ empty, than it is for the H atoms to bind to each $Se_{vac}$ site, separately. Thus, $H_{Se}$ is not likely to form. Besides, the $Se_{vac}$ sites should instead be passivated by strongly bound O, once exposed to ambient conditions [40]. On the other hand, $Se_W$ has been observed in STEM experiments [40]. $Se_W$ does not bind $O_2$, but is the preferred binding site for H (Table II). Thus, H-$Se_W$ is a promising candidate for spin-polarized point defects, and may be engineered by exposing $WSe_2$ ML to atomic H [47].

## C. Charge transition levels and ionization energies

The creation of charge carriers in semiconductors is essential for their use in electronic applications. Unlike $MoS_2$ which is typically n-type, $WSe_2$ ML has p-type or ambipolar transport characterisitics [33-36]. The nature and origin of these charge carriers are fundamental questions in semiconductor physics. Typically, the charge carriers are made available by ionization of defects within the semiconductor [76]. The large surface-to-volume ratio in 2D materials also makes the effects of the substrate and adsorbates particularly important. For example, surface charge transfer effects [77] induced by adsorbing donor or acceptor organic molecules on TMDs can induce doping in the TMD. The nature of the charge carriers is also directly related to the Schottky barrier heights when the electrode comes into contact with the 2D material. Many factors influence the Schottky barrier heights. These include vacuum work function alignment, and Fermi level pinning by metal-induced gap states [78,79], or by defect states [32]. Dielectric screening from the substrate has been found to change the exciton binding energy in $MoS_2$ ML [80]. Likewise, it should be expected that the same substrate screening effect can change the ionization energies of the defects in 2D materials.

Experimentally, it is very challenging to relate measured charge state transition levels to specific defect structures [76]. First principles calculations therefore have a major role to play in studying defect physics, in particular the possibility of inducing doping by ionizing point defects. In bulk materials, the extent of doping caused by point defects is typically analyzed by computing the charge state transition levels and ionization energies (IE), both of which have been defined in Section II. Fig. 13a also presents a schematic explaining these quantities. While the nature and origin of charge carriers in 2D materials are influenced by many factors, we first examine the possibility of ionization of the point defects in $WSe_2$ ML, similar to the way defect-induced doping is studied in the bulk. Previous studies have shown that substrate screening effects on the charge transition levels are quite small in supported $MoS_2$ ML [81]. Thus, to reduce computational complexity, we have computed the charge transition levels in isolated $WSe_2$ ML. The effect of the substrate is then to provide the position of the Fermi level, which can be used to infer the stable charge state of each defect (Fig. 13).

Since the IE are obtained by computing the transition level energies with respect to the band edges, the fact that PBE underestimates the band gap would lead to quantitatively inaccurate IE. Hybrid exchange correlation functionals, such as HSE06 [7], or many-electron GW calculations [82], have in the past been used to obtain more quantitatively accurate charge transition levels. Here, we are more interested in the qualitative features of the IE of various defects. For point defects in $MoS_2$, it has been found that PBE with Grimme's corrections, and HSE06 functionals give qualitatively similar results for the charge transition levels [7]. Thus, we proceed to use PBE-D2 for our calculations. As discussed in Section II, the IE is a quantity that depends on the supercell size, and we have used two different methods to approximate the converged IE in WSe$_2$ ML. In Method 1, we use the asymptotic formula for the IE (equation 5) and obtain the converged IE from the intercepts of linear fits to the data, as shown in Fig. 14 for Se$_{vac}$. The IEs thus computed are used to choose a suitable "special" supercell size for WSe$_2$ ML, to obtain the IE using Method 2 for a larger number of defects. The results are shown in Table IV. Here, we focus on the most stable intrinsic defects (Se$_{vac}$, W$_{vac}$ and Se$_W$), as well as the most stable O- and H-related defects.

### 1. Intrinsic defects

The IE for Se$_{vac}$ are > 1 eV. For W$_{vac}$ and Se$_W$, the donor IE are ~1.5-1.6 eV while the acceptor IE are ~0.8-0.9 eV. These are all deep donors and acceptors and these defects would all be charge neutral in isolated WSe$_2$ ML. For WSe$_2$ ML on graphite, scanning tunneling spectroscopy experiments show that the Fermi level is close to mid-gap, ~1.0 eV from the VBM [58]. Depending on the self-energy corrections for the energy levels, and the alignment of the Fermi level with respect to the VBM for different substrates, it is possible that W$_{vac}$ and Se$_W$ can act as acceptors in supported WSe$_2$ ML. (Note that this assumes that the transition levels are not affected by the substrate.)

### 2. O-related defects

The O-related defects considered here all have large donor and acceptor IE (> 1.5 eV). However, O is well-known to be a very electronegative atom which should behave as an acceptor. Indeed, a Bader charge analysis [83] for the charge neutral states shows that O atoms take 0.9-1.0 electrons from WSe$_2$ ML in O$_{ad}$, O$_{ins}$ and O$_{Se}$, while O in O$_W$ and O-Se$_W$ take 0.8 electrons (Table V). In O$_{Se}$ and O$_W$, the O atom substitutes a Se and W atom, respectively. While O takes 0.98 electrons from its surrounding atoms in O$_{Se}$, Se would also normally take 0.445 electrons (coming from W). Thus, compared to Se, O in O$_{Se}$ takes an additional 0.54 electrons, as indicated in brackets in Table V. On the other hand, O at $W_{vac}$ takes an additional 1.73 electrons compared to W (which normally gives electrons). The $O_2$ molecule has dissociated at $2Se_{vac}$, so the results for $O_2$-$2Se_{vac}$ are similar to those for $O_{Se}$. $O_2$ also takes an average of 0.57 electrons per atom from $Se_{vac}$. On the other hand, $O_2$ takes an average of 0.26 and 0.15 electrons, respectively, when adsorbed on $W_{vac}$ and $Se_W$, significantly less than at other binding sites (Table V). This smaller extent of charge transfer correlates with the fact that $O_2$ does not bind favorably to $W_{vac}$ and $Se_W$ (Table II), thus suggesting that electron transfer to $O_2$ is important for stabilizing the interaction between $O_2$ and the defect sites. Taken together, the Bader charge analysis for O-related defects shows that O atoms in various stable defect geometries do take a considerable amount of electron charge from $WSe_2$. The large acceptor IE computed for these defects may arise from the fact that it is energetically unfavorable to add more electron charge to these O-related defects, which already have some negative charge in the neutral state.

The transfer of electrons to O in the neutral state brings to mind the surface charge transfer doping methods in 2D materials, where donor/acceptor molecules are adsorbed on 2D materials to dope them [77]. However, the concentration of the O-related defects is generally much lower than the density of deposited donor/acceptor molecules. It is not clear if the O atom takes electrons only from its nearest neighbors, leading to local charge rearrangements, or if the O atom also takes electrons from atoms far away; the latter would be required to induce doping in the 2D material. To gain insight into this question, we compute the charge difference $\Delta\rho$, between the amount of Bader charge on a $WSe_2$ unit in a supercell with the defect, and that in perfect $WSe_2$ (equal to 18 valence electrons). The Bader charge on the $WSe_2$ unit is defined as the sum of the charge on W and half of the charges on all six nearest neighboring Se atoms. The quantity $\Delta\rho$ is plotted as a function of the distance r of the $WSe_2$ unit from the defect site, starting from r = 2.6 Å (Fig. 15). Where $\Delta\rho$ is negative, the electron charge in the WSe$_2$ unit there is smaller than that in perfect WSe$_2$, and *vice versa*. This analysis is done for the three most likely O-related defects, O$_{Se}$, O$_{ins}$ and O$_{ad}$ (O$_W$ and O-Se$_W$ have very low densities due to the scarcity of W$_{vac}$ and Se$_W$ defects). Fig. 15a shows that the charge rearrangement at O$_{Se}$ is very localized. Beyond r = 2.6 Å, we do not see much electron charge being depleted from the WSe$_2$ units, and a random distribution of Δρ around 0.0 is obtained. This may be related to the fact that O is at a substitutional site, and O in O$_{Se}$ takes an additional 0.54 electrons only when compared to Se. O$_{ins}$ and O$_{ad}$ seem to be more effective acceptors, with Δρ remaining generally negative for larger r values (Fig. 15b-c). When the graphite substrate is included for O$_{ins}$ and O$_{ad}$, Δρ starts to take random values around 0.0 for r larger than 4.4 and 4.0 Å, respectively (not shown), although the average amount of charge lost per WSe$_2$ unit remains approximately the same with or without the graphite substrate. Thus, it is not clear whether O$_{ins}$ and O$_{ad}$ will be effective acceptors in the presence of the graphite substrate. We note here that graphite does not donate charge to O in these structures. This is in contrast to C$_{60}$F$_{48}$ deposited on WSe$_2$ ML supported on graphite, where it was predicted that graphite contributed a significant amount of charge to C$_{60}$F$_{48}$ [84]. The differing amounts of charge transfer from graphite may be related to the fact that C$_{60}$F$_{48}$ is a much stronger acceptor than O.

### 3. H-related defects

Of all the defects studied here, H$_{ins}$ stands out as having the smallest ionization energy (0.29 eV for the donor IE). This indicates that H$_{ins}$ is an effective donor in WSe$_2$ ML. As discussed above, the interstitial site in H$_{ins}$ is also the most stable binding site for H in perfect WSe$_2$ ML, although the binding energy for the H atom is quite small. This finding indicates that H atoms can be introduced into the WSe$_2$ lattice to induce n-doping, but these H atoms may be removed with annealing unless special precautions are taken. H$_W$ has an acceptor IE of 0.71 eV, slightly lower than that of $W_{vac}$ (0.78 eV). All the other IEs for H-related defects are > 0.9 eV.

## IV. CONCLUSION

In conclusion, we have presented a detailed DFT study of the intrinsic defects in $WSe_2$ ML, supported on graphite, as well as O, H and C-related defects in $WSe_2$ ML. We have considered the interaction of O, $O_2$, H, $H_2$ and C with perfect $WSe_2$ ML as well as with its intrinsic defects. The presence of the graphite substrate does not change the formation energies of the intrinsic defects significantly. When O, $O_2$, H, $H_2$ and C interact with the intrinsic defects, the gap states of the intrinsic defects are modified. In the particular case of O and $O_2$ interacting with Se vacancies and divacancies, the gap states are completely removed. We show that Se vacancies have the lowest formation energy among the intrinsic defects in ML $WSe_2$, and that among all the extrinsic defects studied here, O-related defects ($O_2$ substituting 2Se atoms, O substituting Se and O interstitials in $WSe_2$) are the most stable ones. In reference [40], we further study the optical properties of the most stable defects and comment on the origin of single photon emission from $WSe_2$. H-related impurities are interesting because of the possibility of engineering spin-polarized point defects, such as H bonded to $Se_W$ antisite defects. H also behaves as a donor in its most stable adsorption site in perfect $WSe_2$. All other defects considered here are deep acceptors or donors in isolated $WSe_2$ ML. O defects in their neutral state take electron charge from $WSe_2$, but the charge rearrangement may be too localized for them to be effective acceptors at low densities.

**Acknowledgements:** SYQ acknowledges support from the Singapore NRF, Prime Minister's Office, under its medium-sized centre program, and from grant NRF-NRFF2013-07. Computations were performed on the CA2DM cluster and the National Supercomputing Centre (NSCC) in Singapore.

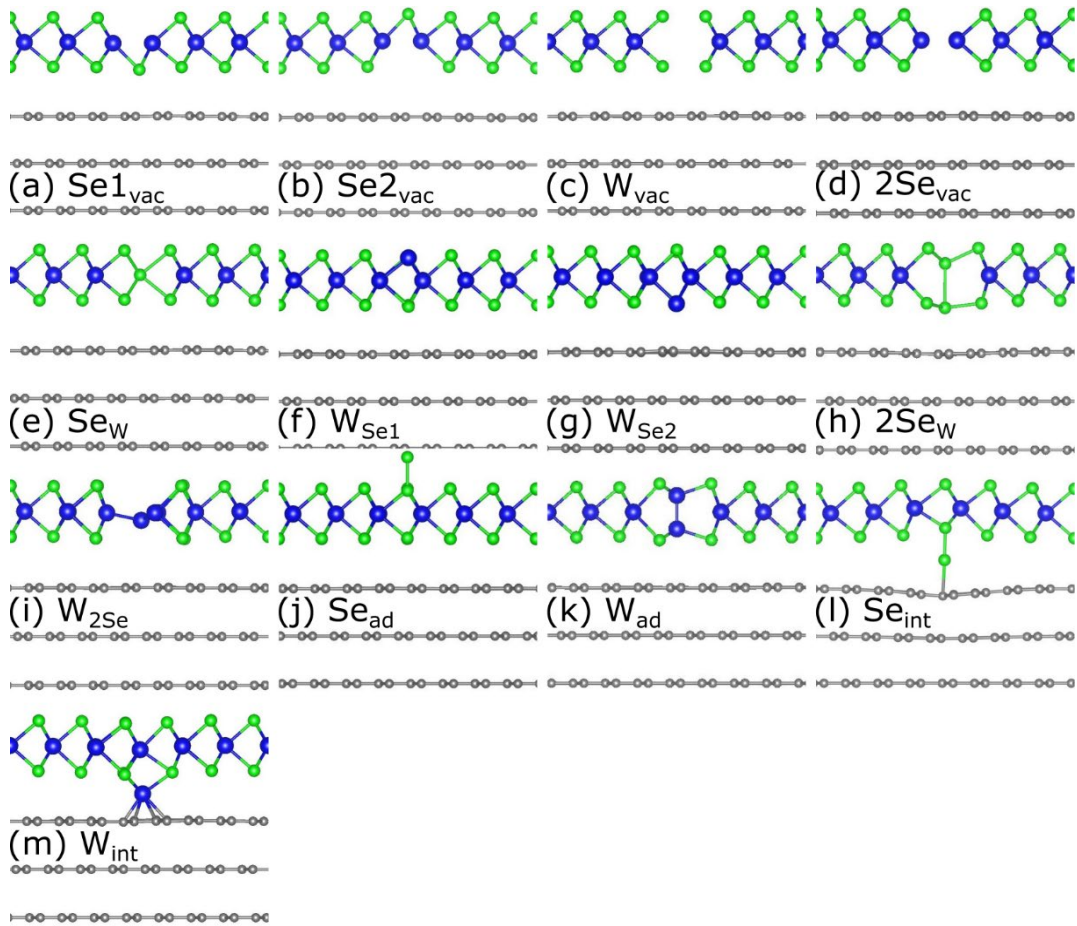

FIG. 1. Side view of the intrinsic point defects in WSe$_2$ ML on graphite. Blue: W, Green: Se, Gray: C. See main text for description of defects.

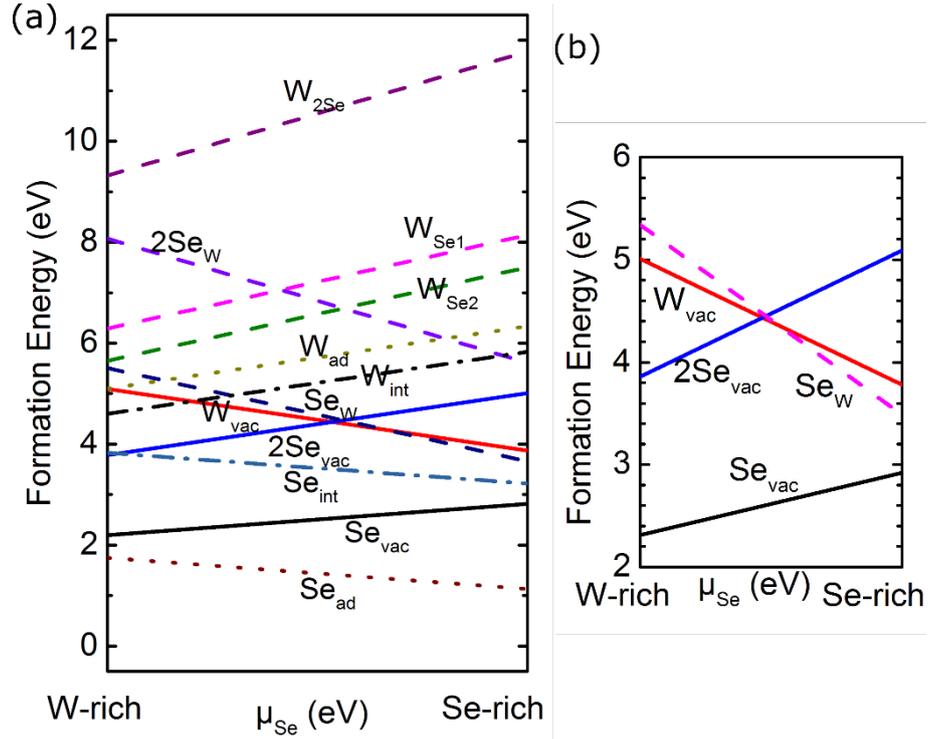

FIG 2. Formation energies of intrinsic defects (a) with and (b) without graphite substrate. Solid lines denote vacancies, dashed lines substitutional (antisite) defects, dotted lines adatoms, and dotted-dashed lines intercalated atoms.

| Defect | W-rich conditions | | Se-rich conditions | |
|---|---|---|---|---|
| | Isolated | Supported | Isolated | Supported |
| $Se_{vac}$ | 2.31 | 2.20 | 2.92 | 2.82 |
| $W_{vac}$ | 5.01 | 5.09 | 3.78 | 3.87 |
| $Se_W$ | 5.34 | 5.51 | 3.49 | 3.66 |
| $2Se_{vac}$ | 3.86 | 3.78 | 5.09 | 5.01 |

Table I. Formation energies in eV of selected intrinsic defects in isolated $WSe_2$ ML and in $WSe_2$ ML supported on graphite.

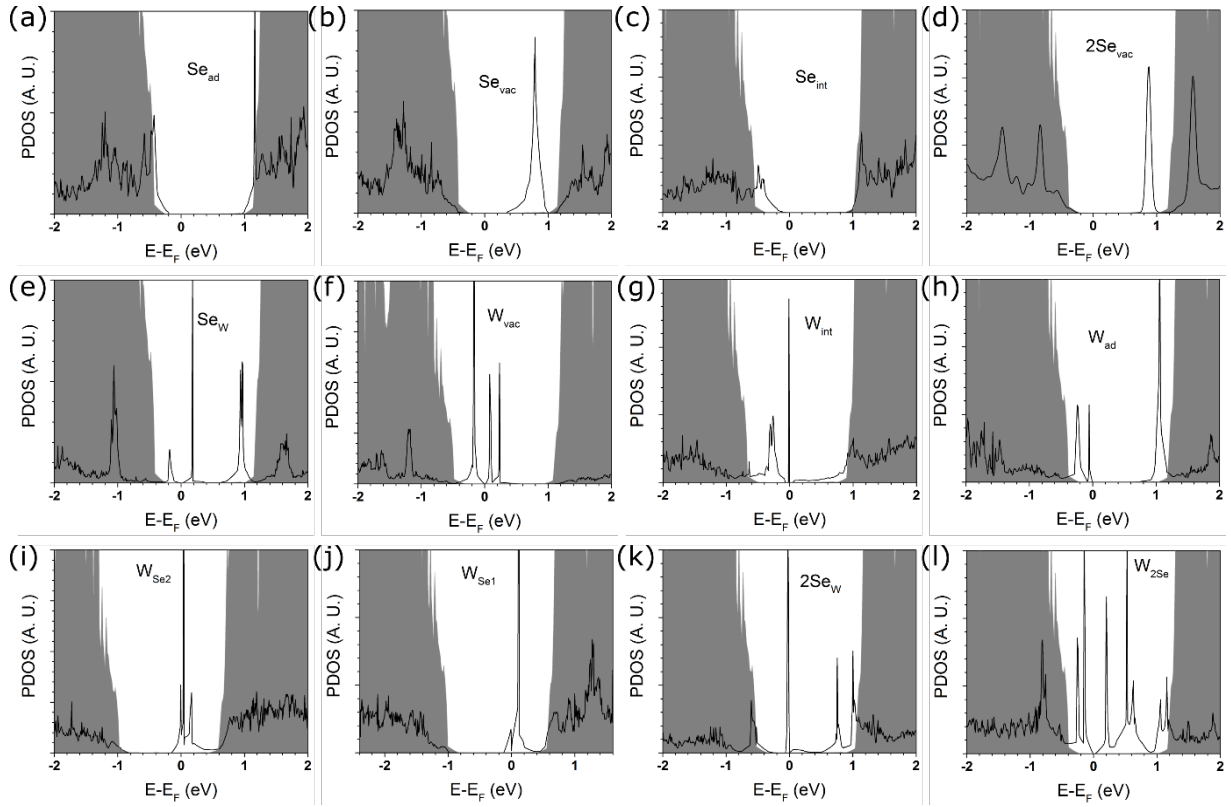

FIG. 3. PDOS on the intrinsic defect sites in WSe$_2$ ML on graphite. Gray shading: PDOS of perfect WSe$_2$ on graphite. The PDOS with and without defects are aligned by the 1s core level of W furthest from the defect site. The Fermi level refers to the Fermi level of WSe$_2$ ML with defects on graphite.

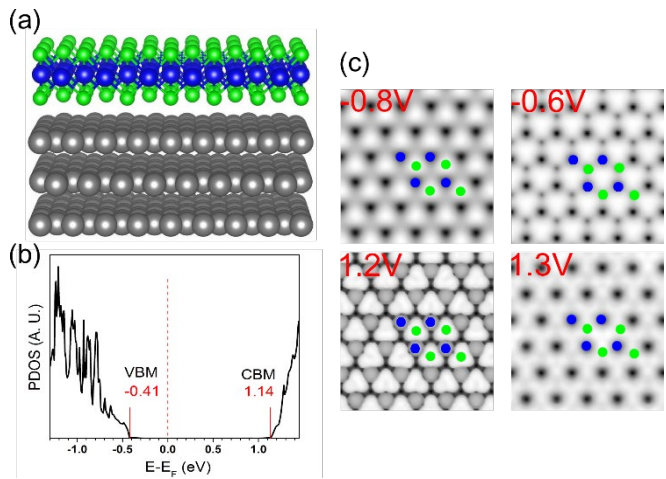

FIG. 4: Perfect WSe$_2$ ML on graphite. a) Atomic structure (Blue: W; Green: Se; Gray: C) b) PDOS on WSe$_2$ c) Simulated STM images at different sample bias voltages with atoms overlain in the images.

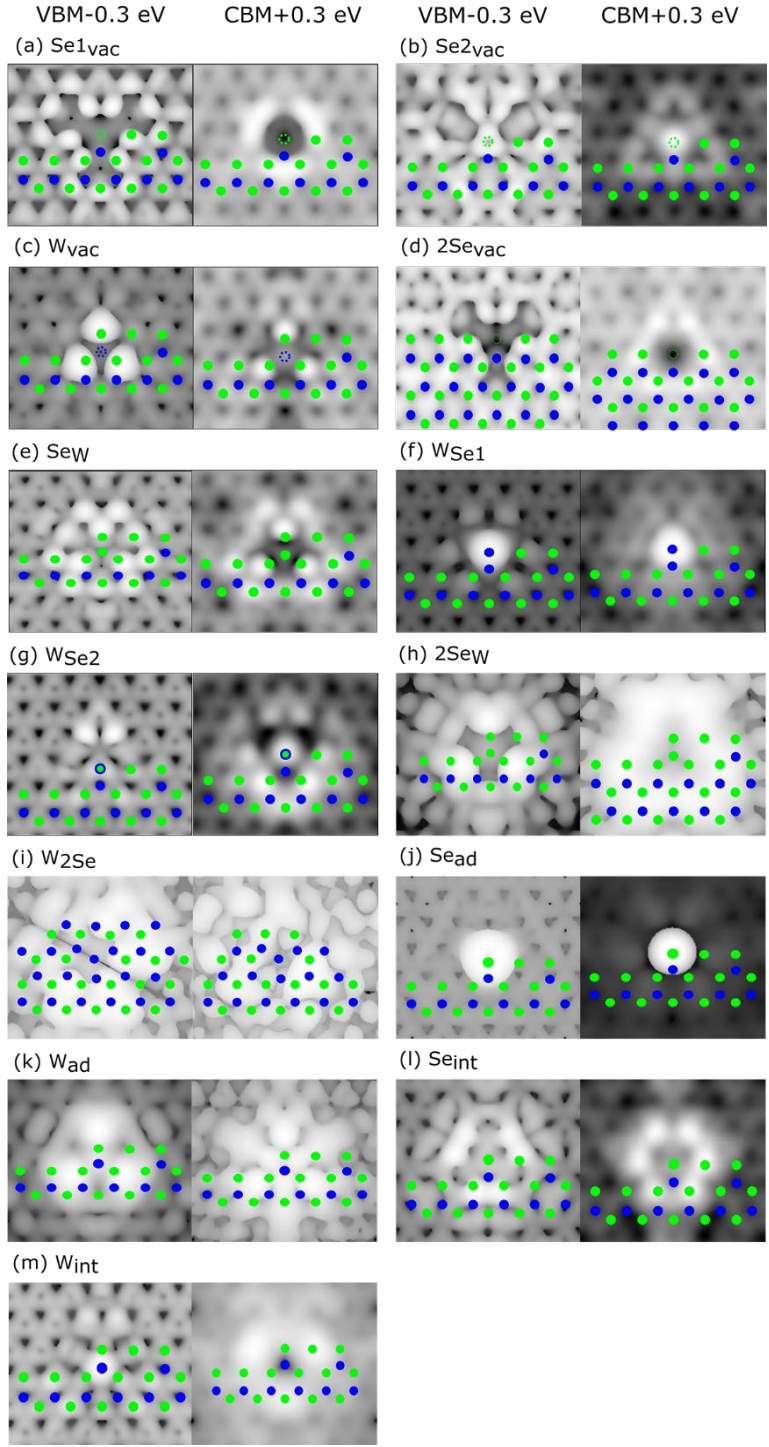

FIG. 5. Simulated STM images of the intrinsic defects in WSe$_2$ ML on graphite. The energy ranges are chosen to probe occupied and unoccupied states that are 0.3 eV away from the band edges. Atomic positions are overlain on the images. Blue: W, Green: Se.

Table II. Formation energies in eV of extrinsic point defects related to O, O$_2$, H, H$_2$ and C. The formation energies $E_{f1}$ and $E_{f2}$ are defined in the main text. Numbers highlighted in bold are the formation energies that are much more negative (more stable) than other extrinsic defects listed here. The chemical potentials of O, H and C are total energy per atom of $O_2$, $H_2$ and graphite, respectively. For numbers in brackets, the chemical potentials of O, H and C are the total energies of gas phase O, H and C atoms, respectively. The O-O or H-H distances are given. For comparison, the bond lengths of gas phase $O_2$ and $H_2$ are 1.23Å and 0.74Å, respectively. Spin polarization: defects in red, italic font have spin-polarized DOS, while all others have no spin polarization.

|  | $O_{ad}$ | $O_{ins}$ | $O_{Se}$ | $O_W$ | O-Se$_W$ | $O_2$-Se$_{vac}$ | *$O_2$-W$_{vac}$* | *$O_2$-Se$_W$* | $O_2$-2Se$_{vac}$ |
|---|---|---|---|---|---|---|---|---|---|
| $E_{f1}$ | **0.22** (-2.36) | **-0.31** (-2.89) | **-1.56** | 2.56 | 3.48 | **0.13** | 4.13 | 4.61 | **-3.47** |
| $E_{f2}$ | - | - | **-4.47** (-7.05) | **-1.22** (-3.80) | -0.01 (-2.59) | **-2.80** | 0.34 | 1.12 | **-8.56** |
| O-O(Å) | - | - | - | - | - | 1.481 | 1.276 | 1.267 | 2.372 |
|  | *$H_{ad}$* | $H_{ins}$ | *$H_{Se}$* | *$H_W$* | *H-Se$_W$* | $H_2$-Se$_{vac}$ | $H_2$-W$_{vac}$ | $H_2$-Se$_W$ | $H_2$-2Se$_{vac}$ |
| $E_{f1}$ | 2.21 (-0.05) | 1.38 (-0.88) | 2.65 | 3.49 | 3.06 | 2.15 | 4.17 | 2.82 | 4.70 |
| $E_{f2}$ | - | - | -0.27 (-2.53) | -0.29 (-2.55) | -0.43 (-2.69) | -0.77 | 0.39 | -0.67 | -0.39 |
| H-H(Å) | - | - | - | - | - | 1.828 | 0.971 | 0.752 | 2.027 |
|  | *$C_{ad}$* | $C_{ins}$ | $C_{Se}$ | $C_W$ | C-Se$_W$ |  |  |  | $C_2$-2Se$_{vac}$ |
| $E_{f1}$ | 6.26 (-1.82) | 3.49 (-4.59) | 3.07 | 4.97 | 6.52 | - | - | - | 5.93 |
| $E_{f2}$ | - | - | 0.15 (-7.93) | 1.19 (-6.89) | 3.03 (-5.05) | - | - | - | 0.84 (-15.33) |

Table III. Effect of different exchange-correlation functionals on the binding energy of $O_2$ at $Se_{vac}$, and the binding energy between O atoms in gas phase $O_2$. All energies are given in eV.

| Exchange-Correlation Functional | Bond dissociation energy in gas phase $O_2$ (experiment: 5.169 eV) | Binding energy between $O_2$ and $Se_{vac}$, relative to energy of $O_2$ (BE1) | Binding energy between $O_2$ and $Se_{vac}$, relative to energy of two O atoms (BE2) | Binding energy between $O_2$ and $Se_{vac}$, relative to energy of two O atoms, computed from BE1 and using the experimental $O_2$ bond dissociation energy of 5.169 eV (BE3) |
|---|---|---|---|---|
| PBE-D2 | 6.04 | 2.80 | 8.84 | 7.97 |
| RPBE | 5.65 | 1.95 | 7.60 | 7.11 |
| RPBE-D3 | 5.65 | 2.20 | 7.85 | 7.37 |
| BEEF-vdW | 5.32 | 1.90 | 7.22 | 7.07 |

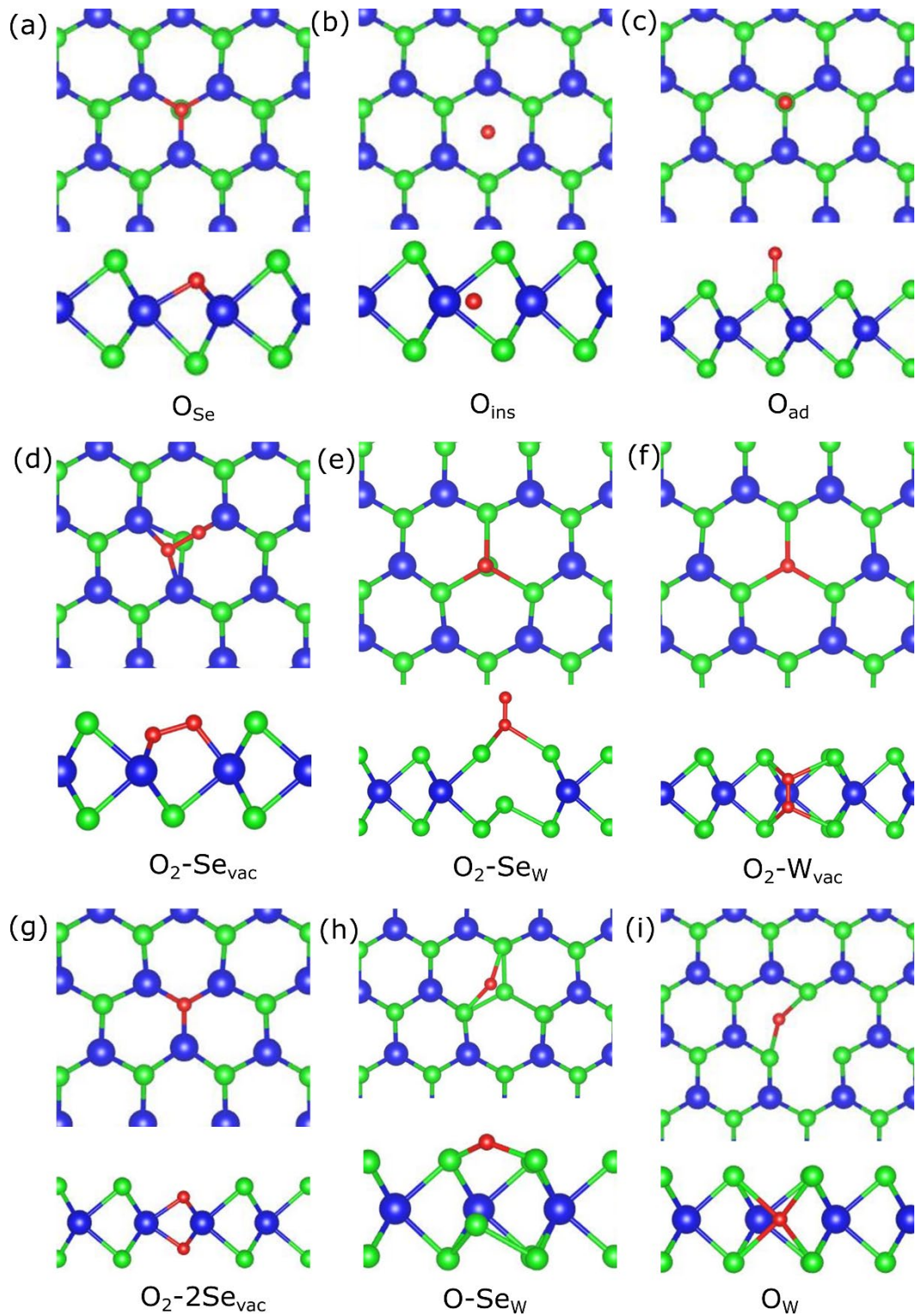

FIG. 6. Top and side view of the atomic structure of O-related defects in WSe$_2$. Blue: W, Green: Se, Red: O.

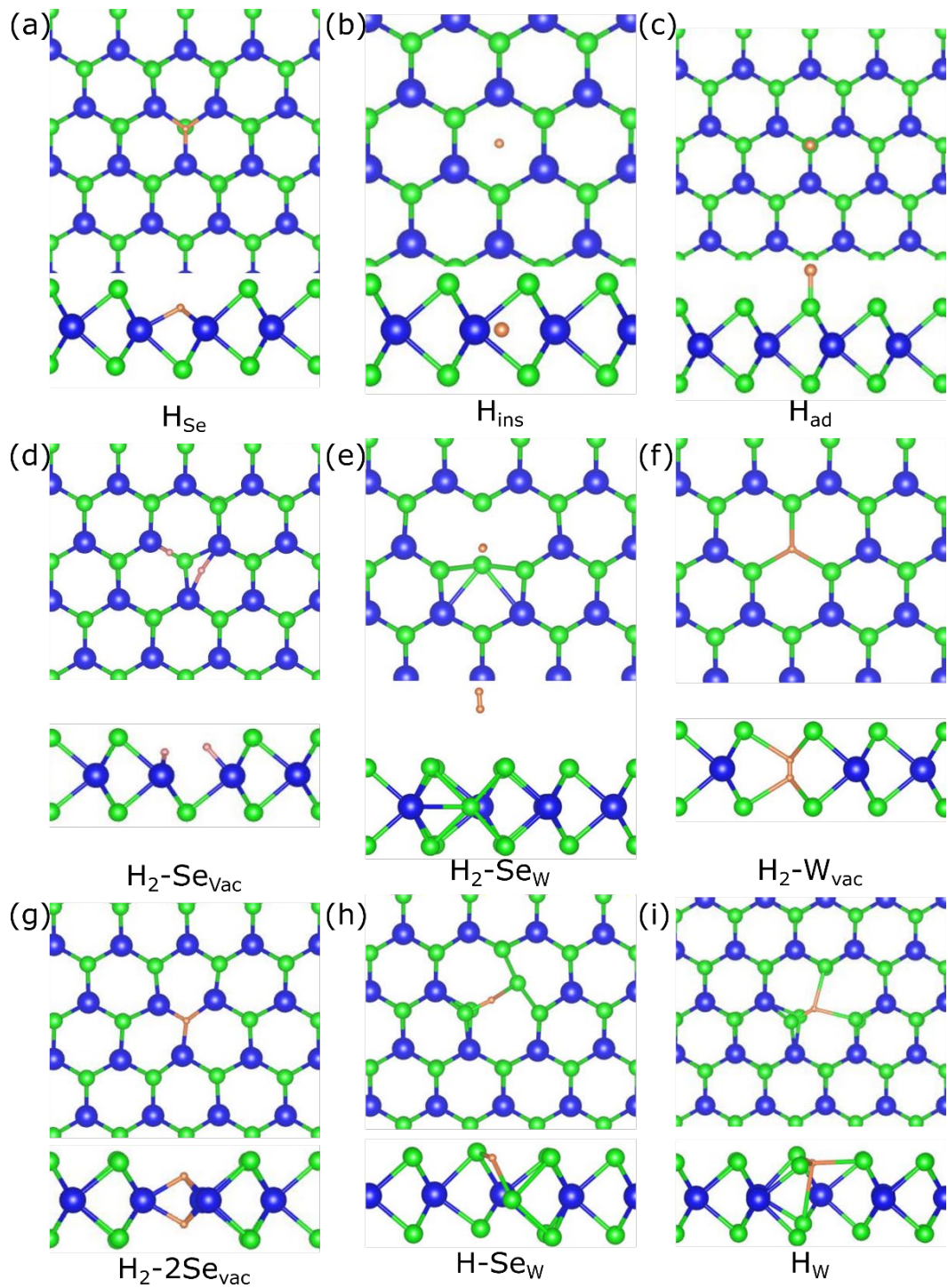

FIG. 7. Top and side view of the atomic structure of H-related defects in WSe$_2$. Blue: W, Green: Se, Orange: H.

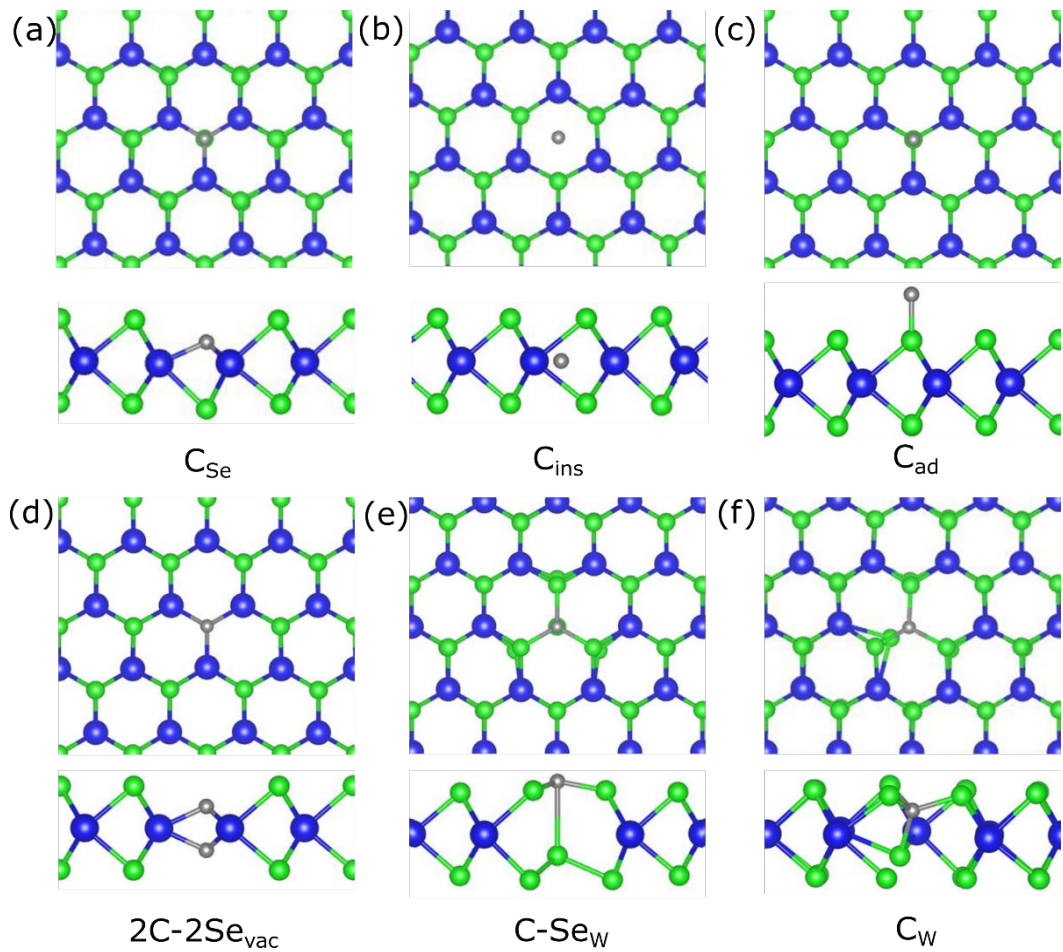

FIG. 8. Top and side view of the atomic structure of C-related defects in $WSe_2$. Blue: W, Green: Se, Gray: C.

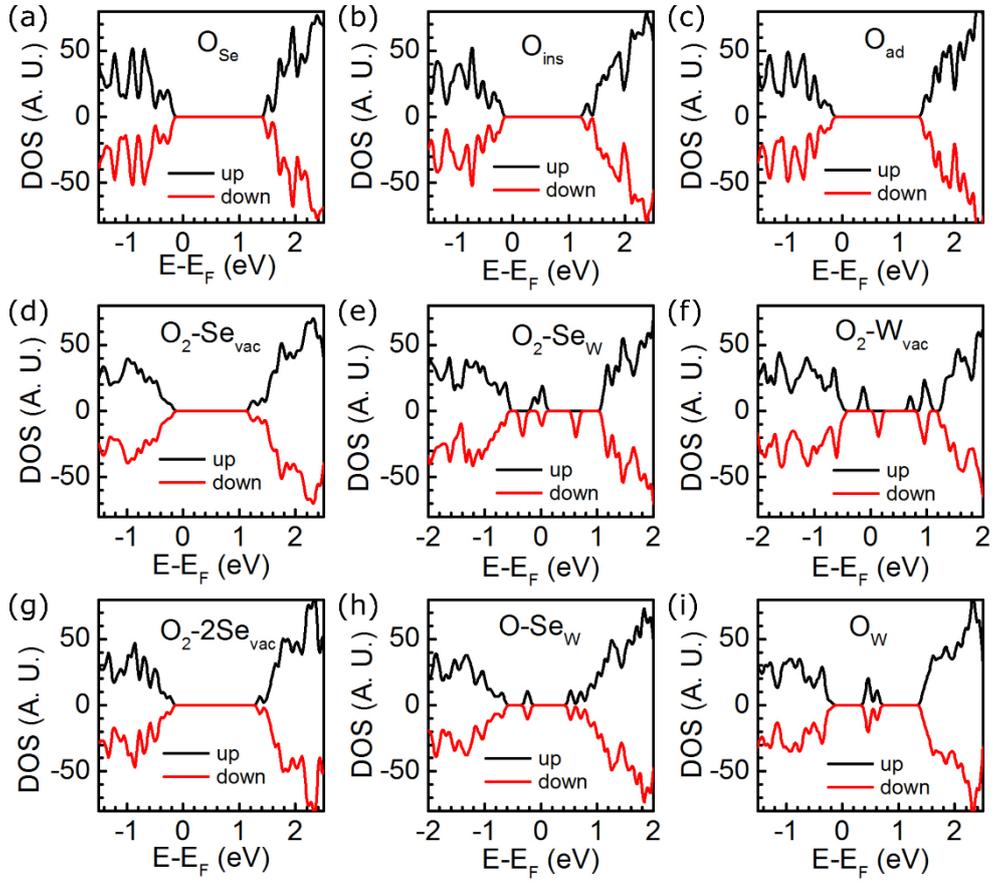

FIG. 9. Spin polarized DOS of O-related defects in WSe$_2$.

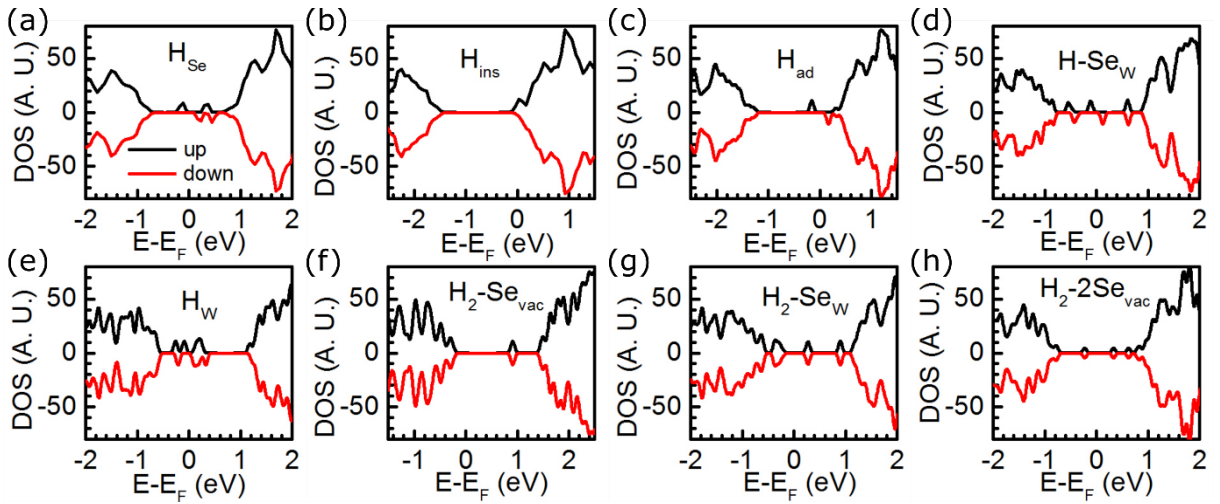

FIG. 10. Spin-polarized DOS of H-related defects in WSe$_2$.
3939

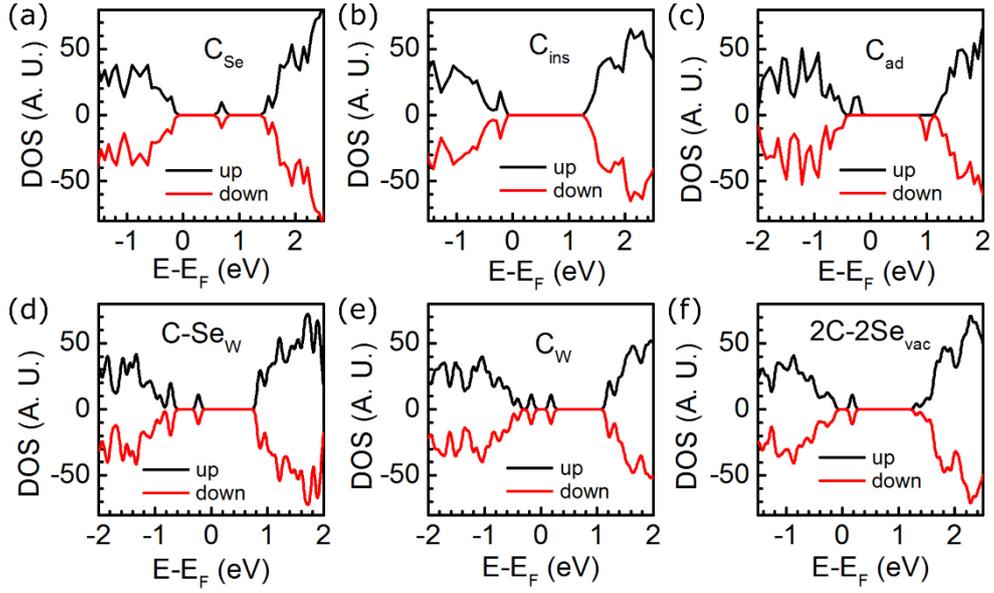

FIG. 11. Spin-polarized DOS of C-related defects in WSe$_2$.

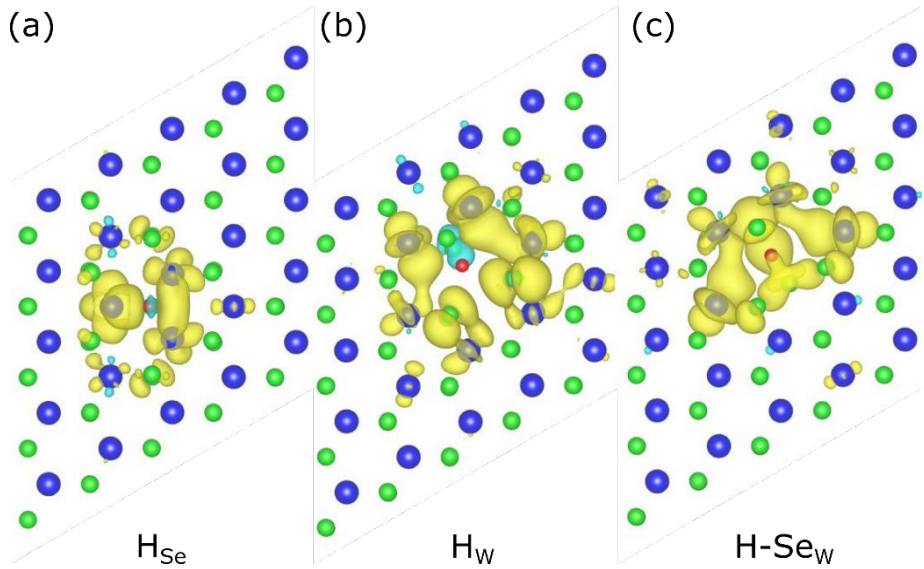

FIG. 12. Isosurface contour plot of the difference between spin up and spin down charge densities, for (a) H$_{Se}$, (b), H$_W$ and (c), H-Se$_W$. Yellow: positive: Blue: negative. Contour values taken to be 2% of the maximum magnitude. H atoms are colored red for clarity. Blue and green circles represent W and Se atoms, respectively.

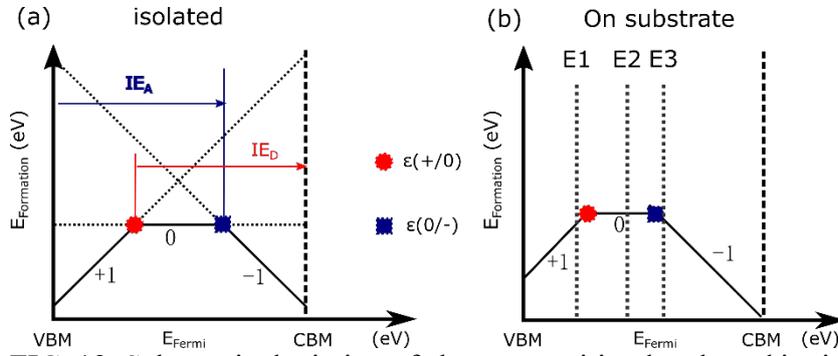

FIG. 13. Schematic depiction of charge transition levels and ionization energies (IE) of defects. (a) Isolated ML. In this paper, the formation energies of the defect in different charged states are computed using the isolated ML. $\varepsilon(+/0)$ and $\varepsilon(0/-)$ are transition levels where the most stable charge state of the defect changes from one charge to another. $\varepsilon(+/0)$ with respect to CBM defines a donor IE ($IE_D$), while $\varepsilon(0/-)$ with respect to VBM defines an acceptor IE ($IE_A$). (b) ML on substrate. When the substrate is included, the Fermi level is determined by the interaction of the ML with the substrate, and this position of the Fermi level in turn dictates the stable charge state of the defect. For example, the defect is in charge state +1, 0 and -1 for $E_{Fermi}$ at E1, E2 and E3, respectively.

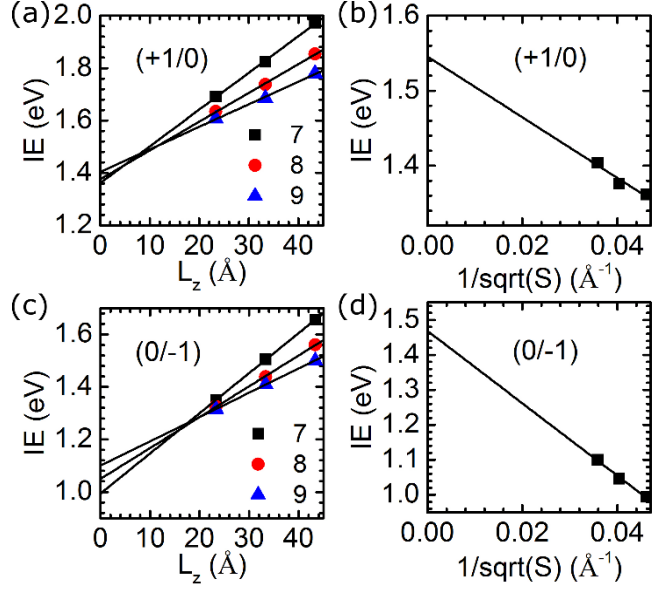

FIG. 14. Plots for computing the ionization energies (IE) of Se$_{vac}$ in WSe$_2$ ML using Method 1 (a), (c) $IE(S, L_z)$ is a linear function of $L_z$ (23.34, 33.34, 43.34 Å) in different lateral dimensions ($S$: 7 × 7, 8 × 8, and 9 × 9) for (+) and (-) charged states, respectively. (b), (d) IE for $L_z = 0$ (vertical intercept in (a,c)) as a function of $\frac{1}{\sqrt{S}}$ for (+) and (-) charged states, respectively. The points fit well to a straight line, and the vertical intercepts in (b) and (d) give the converged IE. The excellent linear fits show that equation 5 (see Section II) holds for Se$_{vac}$.

Table IV. Ionization energies (IE) in eV for the defects obtained using Methods 1 (M1) and 2 (M2) (9×9 supercell with vacuum height of 30 Å).

| IE (eV) | Se$_{vac}$ | W$_{vac}$ | Sew | O$_{Se}$ | O$_{ins}$ | O$_{ad}$ | O$_2$-2Se$_{vac}$ | H$_{Se}$ | H$_{ins}$ | H$_W$ | H-Sew | H$_2$-Sevac | H$_2$-2Sevac |
|---|---|---|---|---|---|---|---|---|---|---|---|---|---|
| donor (M1) | 1.54 | 1.59 | 1.53 | 1.60 | 1.58 | 1.53 | - | - | - | - | - | - | - |
| donor (M2) | 1.69 | 1.50 | 1.47 | 1.69 | 1.69 | 1.68 | 1.69 | 0.97 | 0.29 | 1.31 | 1.04 | 1.68 | 1.21 |
| acceptor (M1) | 1.46 | 0.80 | - | 1.58 | 1.58 | - | - | - | - | - | - | - | - |
| acceptor (M2) | 1.41 | 0.78 | 0.88 | 1.71 | 1.69 | 1.68 | 1.67 | 1.20 | 1.66 | 0.71 | 1.01 | 1.40 | 1.36 |

Table V. Bader charge analysis. Difference between number of electrons centered on the impurity atom, and the number of valence electrons in the neutral impurity atom. A positive number indicates that the impurity atom takes electrons. The numbers in brackets are given for substitutional defects, where the charge normally associated with the missing Se or W atom is subtracted away. For example, O takes 0.98 e⁻ at the $O_{Se}$ site. Compared with what Se would normally take (0.445e⁻), the O takes an extra 0.535 e⁻.

| | $O_{ad}$ | $O_{ins}$ | $O_{Se}$ | $O_W$ | O-Se$_W$ | $O_2$-Se$_{vac}$ | $O_2$-W$_{vac}$ | $O_2$-Se$_W$ | $O_2$-2Se$_{vac}$ |
|---|---|---|---|---|---|---|---|---|---|
| #e⁻ | 0.90 | 1.05 | 0.98 (0.54) | 0.84 (1.73) | 0.83 | 0.62, 0.52 | 0.18, 0.35 | 0.26, 0.05 | 0.97x2 (0.53x2) |
| | $H_{ad}$ | $H_{ins}$ | $H_{Se}$ | $H_W$ | H-Se$_W$ | $H_2$-Se$_{vac}$ | $H_2$-W$_{vac}$ | $H_2$-Se$_W$ | $H_2$-2Se$_{vac}$ |
| #e⁻ | 0.09 | 0.33 | 0.36 (-0.09) | -0.01 (0.88) | 0.05 | 0.32, 0.24 | -0.01, 0.08 | 0.01, 0.00 | 0.34x2 (-0.10x2) |

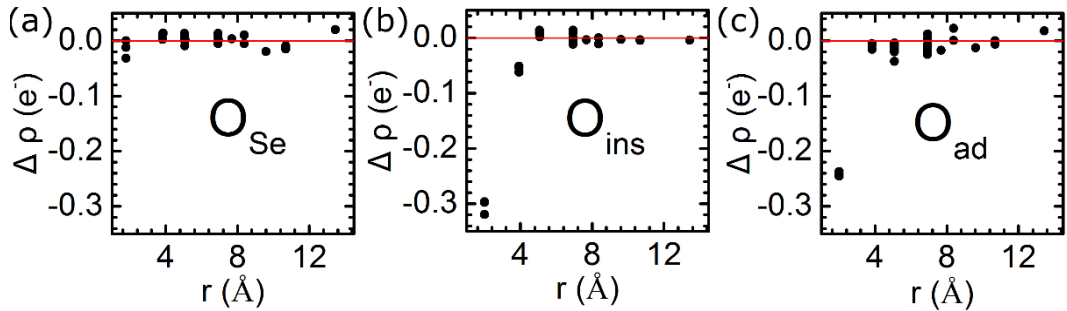

FIG. 15. Bader charge analysis. Charge difference distribution around the defects sites for (a) $O_{Se}$, (b) $O_{ins}$, (c) $O_{ad}$. The charge difference $\Delta\rho$ between the WSe$_2$ units (total Bader charge on one W and 6 nearest neighbor Se atoms (each weighted by 0.5) minus the 18 valence electrons of the perfect WSe$_2$ primitive cell), plotted against r, the distance of the WSe$_2$ units from the defect site. The points for r very close to 0 (where $|\Delta\rho|$ is very large) are absent from the plot.